## GAMMA-RAY BURSTS

# Light on the distant Universe

Jonathan Grindlay

**Observations of a long-lasting γ-ray burst, one that has the brightest optical counterpart yet discovered, challenge theoretical understanding of these bursts but may enhance their usefulness as cosmic probes.**

On a clear night, from one of Earth's increasingly rare dark sites, one can see roughly 3,000 stars with the naked eye. All of these point sources of light are stars within our Milky Way galaxy, and most are closer than about 1500 light years. It is only with the rare catastrophic end of a massive star's life, in a gargantuan explosion resulting from the collapse of the stellar core, that nature extends our visible reach with a supernova.

Possibly one in every thousand supernovae is not 'normal': as the core collapses past the state of a neutron star to a black hole, the spinning disk around the nascent black hole launches a powerful jet that 'drills' its way out of the overlying star[1] and produces an even more extreme blast: a long-duration gamma-ray burst (GRB). These bursts have typical durations of about 3 to 100 seconds, and are followed by fading afterglow emission at longer wavelengths (X-ray, optical infrared and sometimes radio). On page 183 of this issue, Racusin *et al.*[2] report observations of the optically brightest GRB yet seen. The optical emission of this burst, dubbed GRB 080319B, is a hundred times brighter than the previous record holder.

GRB080319B was detected by the Burst Alert Telescope (BAT) onboard NASA's *Swift* satellite on 19 March 2008. Only automated telescopes detected it, but it would have been visible to the naked eye for about 40 seconds — and thus, whoever saw it would have witnessed the most distant astronomical object ever directly seen. Spectra of the optical afterglow measured its redshift as $z = 0.93$ (ref. 3), which corresponds to a light travel time of 7.4 billion years, placing GRB 080319B more than half-way back to the Big Bang and origin of our Universe.

The only 'normal' supernova visible to the naked eye in the past 400 years, SN 1987A, was detected[4] on 24 February 1987. Its optical brightness was comparable to that of GRB080319B, but it occurred a mere 163,000 light years away in our neighbouring satellite galaxy, the Large Magellanic Cloud. How could the similarly bright optical flash of GRB 080319B be in any way connected to the process of stellar death given its approximately $5 \times 10^4$ times larger distance? The answer is 'beaming' — in which, instead of the isotropic, relatively slow emission from a normal supernova over days to months, a large fraction of the total energy of a GRB is collimated into a narrow and highly relativistic jet with bulk outflow velocity very close to the speed of light.

Racusin and colleagues[2] show that the jet in GRB 080319B almost certainly has a two-component structure: a jet approximately 8° across surrounding a narrower (about 0.4°) central core of higher relativistic speed, for which the central jet outflow velocities are within about 5 parts in 10 million of the speed of light. For about 100 seconds, the collimated radiation beam from this jet was an intense beacon illuminating the intervening Universe. It came from a GRB that occurred around three billion years before the Sun and Earth were formed.

X-ray, optical and radio observations of GRBs have shown that their afterglow emission is due to the impact of a beamed jet with the surrounding wind from the pre-supernova star and interstellar medium, and that beaming is directly indicated by the 'jet breaks' in their afterglow lightcurves[5]. Even more convincing evidence for the relativistic expansion of the jet was provided by the radio observations of GRB 970508, which showed[6] that the total GRB energy was about 10 times lower than inferred from a spherical explosion, implying a jet with an opening angle of about 30°. However, until the remarkably complete broadband spectral and temporal coverage of GRB 080319B, it has not been possible to directly constrain the radial structure of the jet.

Observations began before the BAT detection with optical imaging from wide-field telescopes that were already observing another burst, GRB 080319A, which was only 10° away from GRB 080319B on the sky and had gone off only 30 minutes before – a remarkable coincidence given that BAT observes only about two GRBs per week over the full sky. Ultimately, the afterglow from GRB 080319B was observed to fade by eight orders of magnitude in flux over 6 weeks by a world-wide suite of telescopes spanning 11 orders of magnitude in wavelength.

A prediction[2] of the high jet outflow velocities inferred for the central jet is the production of even more luminous, prompt GRB emissions of much higher energy γ-rays. Such emissions would be easily detected by the recently launched Fermi Gamma-ray Space Telescope. But absorption of such high energy γ-rays by the dense optical-ultraviolet photons produced by synchrotron emission in the same internal shock region could attenuate such emissions, despite the small angle scattering in the narrow jet.

The ultra-relativistic core of the jet in GRB 080319B challenges theories of jet formation and, more generally, models of the engine that drives GRBs. If these bursts contain radially structured jets, their usually inferred lower outflow velocities are explained by observations that are almost never exactly aligned with the narrow jet axis. Thus the numbers of GRBs and their production rate, as well as effects on their cosmic surroundings, may be a factor of 10 to 100 greater than previously thought.

This, in turn, bodes well for using the near-infrared spectra of GRBs to measure structure in the Universe at high redshift (Fig. 1). Because long GRBs are almost certainly produced by core collapse of massive stars to rapidly spinning black holes, and because such massive stars are expected to be the first stars formed in the Universe[7], the higher luminosity of jet cores, when aligned, should enable more distant progenitors to be located by observations in the near-infrared. A broader-band, wide-field, X-ray/γ-ray telescope is required that is designed to discover GRBs, and that is more sensitive and has higher spatial/spectral resolution than Swift/BAT, to enable prompt near-infrared spectroscopy to be carried out with an optical-infrared telescope on board. That would allow simultaneous GRB identification and redshift determination, as well as determination of the intervening cosmic structure. The Energetic X-ray Imaging Survey Telescope, *EXIST*, is a combined wide-field γ-ray imaging and optical/near-infrared imaging spectroscopy telescope[8] designed to discover and map the very first cosmic stellar black holes, and is under study by NASA for a proposed mission in the coming decade.

Jonathan Grindlay is at the Harvard-Smithsonian Center for Astrophysics, Cambridge, Massachusetts 02138, USA.
e-mail: josh@head.cfa.harvard.edu

# Figure and Caption

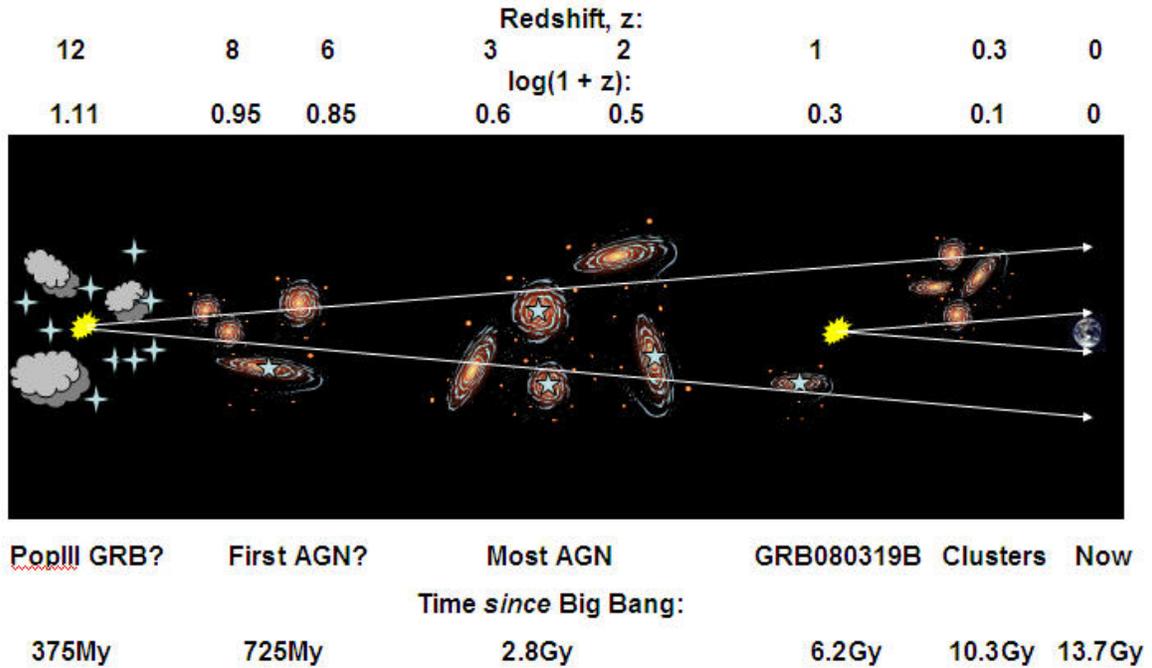

**Figure 1: Timeline of the Universe since formation of first stars.** Dense clouds of gas collapse into the first (Pop III) massive stars and probably produce the first GRBs (ref 7 and references therein). GRBs can then precede formation of the first galaxies, which in turn precede that of active galactic nuclei (AGN) powered by supermassive black holes. Thus, GRBs could probe the first structures and galaxies to emerge after the 'dark ages' of the Universe. The narrow beaming of GRBs, better defined by GRB080319B (not to scale), makes them the most luminous back-lights for mapping the far-distant visible Universe.